# SCIENCE ON THE LUNAR SURFACE FACILITATED BY LOW LATENCY TELEROBOTICS FROM A LUNAR ORBITING PLATFORM-GATEWAY

**Jack O. Burns, Benjamin Mellinkoff, and Matthew Spydell**
University of Colorado Boulder
jack.burns@colorado.edu, Benjamin.mellinkoff@colorado.edu, matthew.spydell@colorado.edu

**Terrence Fong**
NASA Ames Research Center
terry.fong@nasa.gov

**David A. Kring**
Lunar and Planetary Institute
kring@lpi.usra.edu

**William D. Pratt, Timothy Cichan, and Christine M. Edwards**
Lockheed Martin Space Systems Company
william.d.pratt@lmco.com, timothy.cichan@lmco.com, christine.m.edwards@lmco.com

**Abstract**: NASA and ESA are preparing a series of human exploration missions using the four-person Orion crew vehicle, launched by NASA's Space Launch System, and a Lunar Orbiting Platform-Gateway (LOP-G) that enable long duration (>30 days) operations in cis-lunar space. This will provide an opportunity for science and exploration from the lunar surface facilitated by low latency surface telerobotics. We describe two precursor experiments, using the International Space Station (ISS) and a student-built teleoperated rover, which are laying the groundwork for remote operation of rovers on the Moon by astronauts aboard the LOP-G. Such missions will open the lunar far side, among other sites, for exploration and scientific exploration. We describe examples of two high-priority, lunar science missions that can be conducted using low latency surface telerobotics including an astronaut-assisted far side sample return and the deployment/construction of a low frequency radio telescope array to observe the first stars and galaxies (Cosmic Dawn). The lessons learned from these lunar operations will feed-forward to future low latency telepresence missions on Mars.

**Declarations of interest:** None

## 1. INTRODUCTION

Within the next few years, NASA's Space Launch System (SLS) is scheduled to dispatch the Orion crew vehicle including ESA's European Service Module into lunar orbit where it will begin human exploration beyond Low Earth Orbit (LEO) for the first time in half a century. A series of exploration missions are planned for cis-lunar space to evaluate crew health and spacecraft performance in deep space in order to prepare for long duration missions to Mars. In addition, NASA is planning to place a habitable spacecraft, called the *Lunar Orbiting Platform-Gateway* (LOP-G or Gateway), at a location such as the Earth-Moon L2 Lagrange Point above the lunar far side [1] or within a Near-Rectilinear Halo Orbit (NRHO) [2]. The LOP-G will include a propulsion system that would support transfer between lunar orbits. Orion would dock at the Gateway and permit extended stays of one or two months. In the future, the LOP-G could also provide a docking station for a reusable lander (possibly provided by international or commercial partners) which would then offer astronaut access to both the near and far sides of the Moon's surface.

During the early phases of development of infrastructure near the Moon, there is an exciting opportunity to begin a new era of science and exploration by employing low latency surface telerobotics, i.e., having astronauts in orbit remotely operate rovers, or other robots, on the lunar surface. Forefront science (e.g., as described in the U.S. National Academies NRC Decadal Surveys) can be conducted from the lunar far side using low latency surface telerobotics as a harbinger for returning humans to the Moon. In this paper, we discuss two possible examples of science missions from the unexplored lunar far side that could be facilitated by surface telepresence. First, a planetary rover remotely operated by astronauts on the LOP-G could enable the collection and return of multiple rock samples from the Moon's South Pole-Aitken (SPA) basin, as recommended by the NRC Planetary Sciences Decadal Survey [3] and the NRC-2007 [4] report. Second, astronauts on the LOP-G could deploy/construct a low frequency radio telescope array to observe the redshifted 21-cm signal from neutral hydrogen originating from structure within the intergalactic medium surrounding the first stars and galaxies; these observations help to fulfill recommendations from the NRC Astrophysics Decadal Survey [5] and NASA's Astrophysics Roadmap (Cosmic Dawn Mapper [6]).

We begin this paper in Section 2 with an elaboration of the capabilities of the Lunar Orbiting Platform-Gateway and Orion in cis-lunar space that will enable robotic missions on the lunar surface with particular emphasis on the lunar far side. In Section 3, we describe experiments using the ISS and a student-built teleoperated rover that are helping to define requirements for cis-lunar telepresence operations. In Section 4, science from the lunar far side expedited by low latency surface telerobotics is described including astronaut-assisted sample return and low radio frequency observations of Cosmic Dawn. A summary and conclusions are given in Section 5.

## 2. EXPLORATION WITH ORION AND THE LUNAR ORBITING PLATFORM-GATEWAY

### 2.1 The Gateway as Supporting Infrastructure for Lunar Low Latency Surface Telerobotics

NASA has outlined a phased approach to expand human presence deeper into the solar system, starting with the Moon. Phase 1 of this plan begins in the 2020s, with missions and assembly of the Gateway in cislunar space. The LOP-G will be a space platform comprised of several elements which provide key capabilities for cislunar exploration; a habitat where the crew would live and work, a self-sufficient power and propulsion bus with docking capability, an extra-vehicular activity (EVA) element with an airlock for spacewalks and storage, and a cargo/logistics pod for supplies and trash disposal. Lockheed Martin (LM) is one of the companies currently working on a design of the Gateway through the NASA funded NextSTEP Habitat program. Figure 1

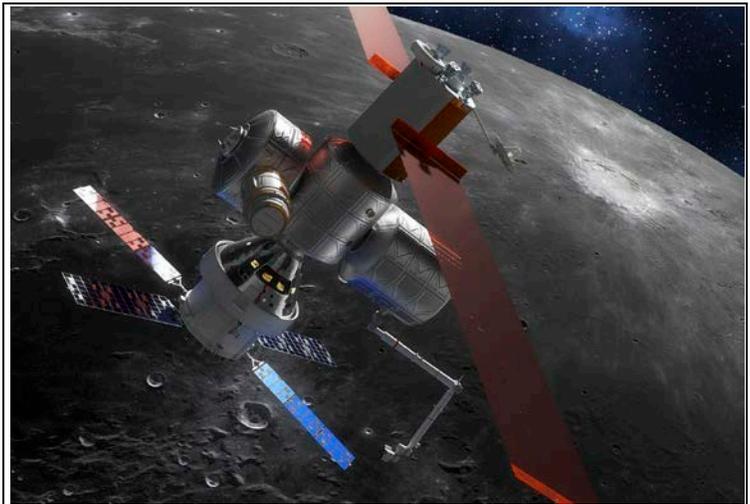

**Figure 1.** Lockheed Martin's Gateway concept supports lunar science objectives and serves as a state-of-the-art telerobotics platform.

illustrates a possible concept for the Lunar Orbiting Platform-Gateway.

The Gateway will be used to achieve lunar science objectives while simultaneously laying the groundwork for future deep space missions. It will operate in two overall mission modes, crewed and uncrewed, which each contribute different strengths to lunar science support. During crewed missions, astronauts are on-board the LOP-G for up to 30 days initially each year. The LOP-G's capability to support much longer durations will increase with each mission (beyond 60 days). During this period of time, the crew will be able to perform scientific experiments in the LOP-G, provide waypoint services for lunar landers and sample returns, and also support other lunar science from cis-lunar space involving low latency surface telerobotics. The low-latency commanding discussed in this paper is one of the strengths of crewed missions that can be provided by humans-in-the-loop onboard the Gateway. When the astronauts are physically located closer to the lunar surface, the latency can decrease from multiple seconds to subseconds, enabling commanding that is closer to real-time and allowing for operations that are more streamlined than remote robotics [7].

When uncrewed, the Gateway will be operated like other robotic planetary missions with remote flight control teams commanding the spacecraft from Earth. One of the Gateway's services will be to provide communication relay for robotic spacecraft (landers, planetary rovers, etc.), which will be especially critical for any future far side surface assets. The lunar far side is of scientific interest, but currently no communications architecture exists there to support scientific missions. The LOP-G can act as a communications relay to any surface or orbital missions, including international missions, and can transmit the data back to scientists on Earth via either the High Gain Antenna (HGA) or the optical communication terminal. The science and video data collected by the LOP-G is most efficiently downlinked using an optical communication system. Optical communication enables a significant increase to downlink bandwidth capability compared to traditional radio frequency (RF) communication. For example, NASA's Lunar Laser Communication Demonstration (LLCD) exhibited a record-breaking Moon to Earth download rate of 622 Mbps [8]. The Gateway will also have similar capabilities as the Mars orbiters, including Mars Reconnaissance Orbiter, Mars Odyssey, and MAVEN, which provide communications between the Mars rovers and Earth [9]. With the currently planned Near Rectilinear Halo Orbit (NRHO), the Gateway will be in a position to provide more than 60% communication coverage for a large portion of the lunar far side, which is greater than the relay support coverage provided by the current Mars fleet [10]. In addition, the LOP-G could provide a range of other surface mission support services, including remote sensing and imaging, "orbital computing" (high-performance computation and data storage similar to terrestrial "cloud computing"), positioning, and timing.

For both crewed and uncrewed lunar mission support, the Gateway communications architecture is designed to exchange commands and telemetry via space to ground, space to space, and space to lunar surface links. The LOP-G communications system has deep space heritage from planetary spacecraft and Orion to meet the various communication needs and utilizes channels across the frequency spectrum including X-band, S-band and optical communication. While the LOP-G is in lunar orbit without Orion, commands and telemetry will be exchanged over an X-band link with the Deep Space Network (DSN) via a 2 meter HGA. This communication link will then be used for remote operations of the Gateway as an uncrewed robotic spacecraft. It will also be the path for downlinking the relay data from other robotic missions. During rendezvous and proximity operations, the LOP-G will communicate with Orion over S-Band. The LOP-G will leverage S-Band components developed for Orion to ensure compatibility and affordability. Once docked, data will be exchanged between LOP-G and Orion via a hardline connection. Information from Orion can be downlinked to the DSN via the LOP-G's HGA; and vice versa, information from the LOP-G can be downlinked via Orion's Phased Array Antennas that are part of its S-Band system.

There are several key lunar far side science objectives discussed in this paper, which have been identified by the planetary science and astrophysics communities as top priorities and can be supported through use of the Gateway. For example, the relay capability will allow teleoperation from Earth of lunar robots to perform surface imaging, in-situ chemical and isotopic analyses, and sample collecting opportunities in scientifically-rich sites such as the Schrödinger basin [11]. Sample collection from Schrödinger and other locations is required to resolve the questions such as in the NRC-2007 report outlining the *Scientific Context for Exploration of the Moon*. The LOP-G could also provide a platform for the continuous observation of impacts onto the lunar surface [12]. This capability would provide an enhanced assessment of impact hazards for lunar surface operations and identify fresh surface excavation sites suitable for geologic and in-situ resource utilization studies.

## 2.2 The Orion Crew Vehicle

As the next step in the progression of human spaceflight into deep space, the Gateway provides the unique opportunity to merge the strengths of human and robotic spaceflight missions. As discussed in the previous section, the crewed and uncrewed phases provide different strengths in lunar science support. In not just lunar science support but also in overall operations, the Gateway could capture a "bliss point" of the human-machine interface by finding and implementing that ideal combination of crewed and uncrewed capabilities. This bliss point can be achieved in part through a symbiotic relationship between the Gateway and the Orion crew vehicle. Aspects of the avionics, crew interface, life support, power, communication, and navigation systems on Orion, which are highlighted in Figure 2, can be utilized to minimize duplication and rework on the Gateway while providing a safe environment for astronauts to live and work. These advanced subsystems on Orion can be used to increase safety and affordability of the Gateway by minimizing duplication and rework and providing redundant and proven systems for a safe environment where astronauts will live and work.

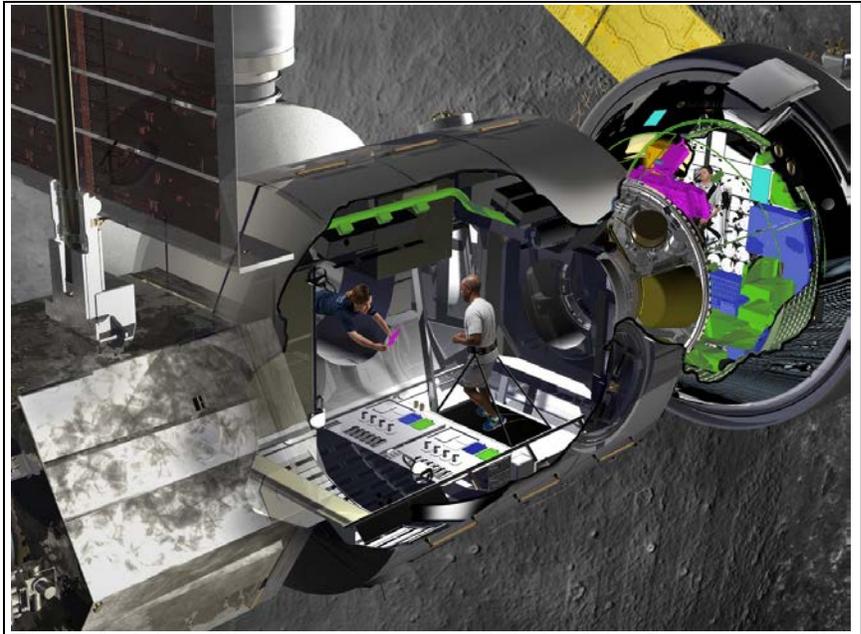

**Figure 2.** Orion will serve as the command deck of the Gateway, providing advanced functionality needed to support human presence. Lockheed Martin concept.

When attached to the Gateway, the Orion Command and Control systems will cover critical functionality: functional availability and functional safety. This connection then enables the Gateway to meet key performance requirements with less complexity and lower-cost systems of its own. There are four redundant Flight Computer Modules (FCMs) within the two Vehicle Management Computers (VMCs) on Orion. The FCMs provide a high integrity platform to house software applications. The FCMs have sufficient processing power to perform command and control of Orion as well as the LOP-G. Utilizing Orion as the command deck when docked with the Gateway enables a more streamlined approach to the avionics on the LOP-G. The LOP-G command and data handling can be more akin to a deep space planetary mission because it relies on the reliability and availability built into Orion [13].

In the unlikely event that something goes amiss with the primary flight computers on Orion, a dissimilar processing platform with dissimilar flight software is hosted on the Vision Processing Unit (VPU). The VPU provides a hot backup function to the redundant FCMs during critical phases of flight. This capability can also be utilized by astronauts aboard the LOP-G should emergencies arise in cis-lunar space.

Orion employs a wireless communication system to interface with cameras used to monitor critical events and crew activities. This system is capable of sending commands and receiving telemetry from end systems and is connected to a utility network that interfaces with the Orion Onboard Data Network (ODN). With the use of portable tablets and the Orion wireless communication system, the crew has flexibility to be in any area of the combined Orion/Gateway and have insight into the critical systems of the cis-lunar station while having the ability to act on any urgent caution, warning or emergency alerts.

The Orion Displays and Control equipment is the crew interface to its subsystems. The Display Units (DUs) utilize a variety of Display Formats to provide data to the crew for awareness and action when necessary. The Display Format Software Architecture enables streamlined addition of new formats via the Generic Display Engine or for more complex formats via a library of reusable and common graphical elements. This library of graphical elements can be leveraged to facilitate development of unique Formats for the LOP-G which can be displayed on the Orion DUs or the supplemental wireless tablet. Electronic Procedures have been developed for Orion that allow direct interaction with the Display Formats enabling reduced workload on the crew. This same methodology can be employed on the LOP-G, providing the crew more time to accomplish more mission objectives and increase science return.

There is a symbiotic relationship between Orion's Environmental Control and Life Support Systems (ECLSS) and what is needed on the Gateway. Utilizing the waste management system and galley with water dispenser on Orion prevents the need to duplicate those systems on the LOP-G. Orion's regenerative pressure swing amine beds can simultaneously remove $CO_2$ and humidity during docked operations, which reduces the load required for the LOP-G to handle. Additionally, Orion utilizes a regenerative Phase Change Material (PCM) heat exchanger to accommodate peaks of high thermal loads rather than expendable consumables, which may reduce complexity of the LOP-G/habitat system.

The Orion power system is capable of generating and supplying more power than is required for its on-orbit operations and surplus power can be shared with the LOP-G to supplement science experiments performed by the crew. The four Orion solar arrays generate about 11 kW of power and extend 62 feet when fully deployed. Orion's batteries use small cell packaging technology to ensure crew safety when providing 120 V power to the many systems on Orion and this technology can also be leveraged to ensure a safe environment while the crew is onboard the Gateway.

Because of this integration with Orion and this symbiosis between human and robotic capabilities, a large portion of the workload for maintaining the Gateway will be shared across multiple automated and remote systems. The more-complex Orion systems will be replaced or refurbished on Earth between missions, and the simpler Gateway systems will mostly be maintained remotely and robotically with minimal astronaut servicing. As a result, the astronauts will have more time to spend on lunar science support. And when the Gateway is uncrewed, its systems will be sufficient for operating as a remote robotic spacecraft, as discussed in the previous section. Achieving the bliss point of human and robotic spaceflight will then enable the kind of lunar science missions discussed below.

## 3. LOW LATENCY SURFACE TELEROBOTICS EXPERIMENTS

Since the early 1960s, humans have been exploring space through an on-going series of missions. Many of these missions have involved short-duration, orbital flights (the Space Shuttle, Soyuz, etc). Other orbital missions have focused on long-duration space stations (Mir, Skylab, and the ISS). Beyond LEO, the Apollo missions orbited and landed humans on the Moon.

In planning for future human space exploration, numerous NASA and international study teams have hypothesized that astronauts can efficiently remotely-operate surface robots from a flight vehicle [14][15][16][17][18]. This concept of operations is seen as a cost-effective method for performing surface EVA activities. Moreover, it is believed that such "low latency surface telerobotics" can enhance and extend human capabilities, enabling astronauts to be telepresent on planetary surfaces in a highly productive manner.

Many assumptions have been made regarding surface telerobotics, including technology maturity, technology gaps, and operational risks. Although many related terrestrial systems exist (e.g., unmanned aerial vehicles), integrating telerobots into human space exploration raises several important questions. What system configurations are effective? Which modes of operation and control are most appropriate? When is it appropriate to rely (or not) on telerobots? How does communications availability, bandwidth, and latency impact productivity? When should rovers operate in autonomous modes versus under manual control (direct human teleoperation)?

To examine these assumptions and to answer these questions, we developed a series of tests (including use of the ISS) to: (1) demonstrate interactive crew control of a mobile surface telerobot in the presence of a short communications delay, (2) characterize a concept of operations and (3) characterize system utilization and operator work for a single astronaut remotely operating a planetary rover with limited support from ground control [21].

These tests were motivated by the fact that although a significant amount of ground-based laboratory and analogue mission testing had previously been performed, no crew-controlled surface telerobotics system had been flight tested in a fully operational manner and characterized using detailed performance metrics (workload, situation awareness, etc).

## 3.1. International Space Station Experiment

During summer 2013, we conducted initial testing of the "Orion/LOP-G L2 Far side" mission concept using the ISS in LEO as a proxy for the Gateway in cis-lunar space [19]. Over the course of ISS Expedition 36, astronauts Chris Cassidy, Luca Parmitano, and Karen Nyberg on the ISS remotely operated NASA's "K10" planetary rover in the "Roverscape" analogue lunar terrain located at the NASA Ames Research Center. The astronauts used a Space Station Computer (Lenovo Thinkpad laptop), supervisory control (command sequencing with interactive monitoring), teleoperation (discrete commanding), and Ku-band satellite communications to remotely operate K10 for a combined total of 11 hours.

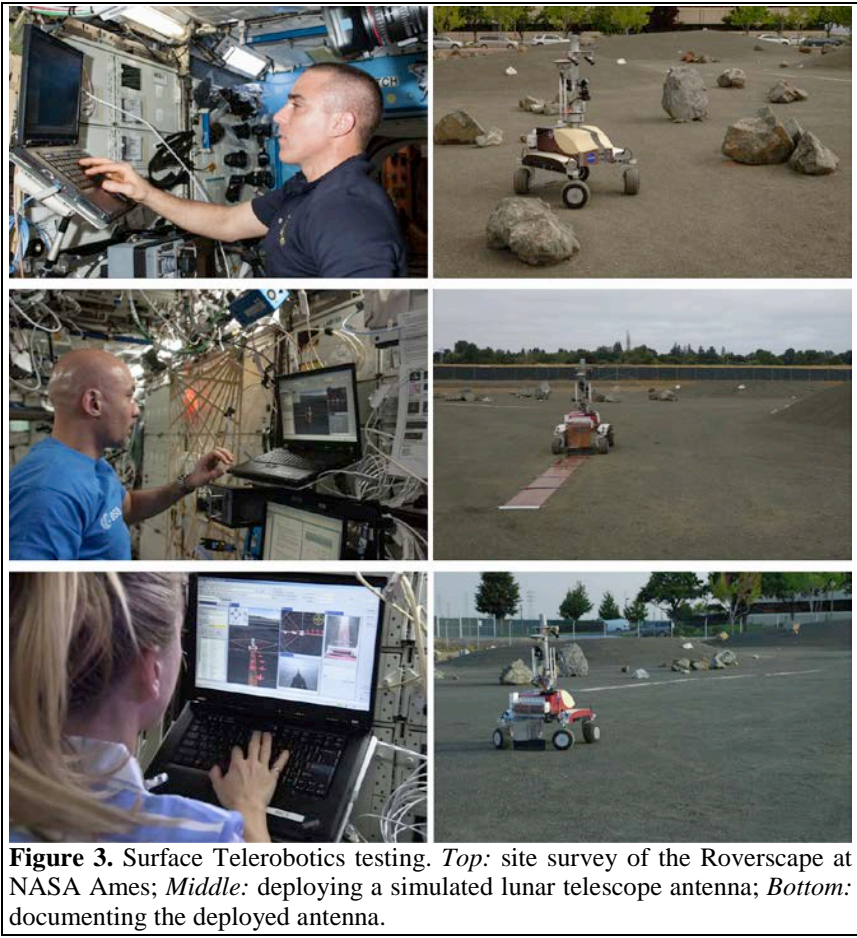

**Figure 3.** Surface Telerobotics testing. *Top:* site survey of the Roverscape at NASA Ames; *Middle:* deploying a simulated lunar telescope antenna; *Bottom:* documenting the deployed antenna.

The testing was designed to simulate four mission phases: (1) pre-mission planning, (2) site survey, (3) radio telescope deployment, and (4) telescope inspection. We performed the pre-mission planning phase using satellite imagery of the test site at a resolution comparable to what is currently available for the Moon and a derived terrain model to select a nominal site for deployment. In addition, the planning team created a set of rover task sequences to survey the site, looking for hazards and obstacles.

Following pre-mission planning, we carried out the three other mission phases during three crew sessions. These sessions were designed to be increasingly difficult (in terms of task sequence complexity, number of contingencies/difficulties encountered, etc.) in order to study the impact of difficulty on crew workload and situation awareness. Since none of the astronauts had prior experience with K10 or the operator interface, each session included an hour of "just-in-time" training. After training, each astronaut then worked with K10 for approximately two hours.

On June 17, 2013 (Session 1), NASA Astronaut Chris Cassidy remotely operated K10 to survey the Roverscape site (Figure 3, top). The survey data collected with K10 enabled assessment of site characteristics, including obstacles (e.g., large rocks), slopes, and other terrain features. Surface-level survey complements remote sensing data acquired from orbit by providing measurements at resolutions and from viewpoints not achievable from orbit. In particular, K10 provided close-up, oblique views of the locations planned for telescope deployment.

On July 26, 2013 (Session 2), ESA Astronaut Luca Parmitano used K10 to deploy three "arms" of a simulated radio telescope array (Figure 3, middle). Parmitano first executed each task sequence with the deployment device disabled, to verify that the sequence is feasible. He then commanded K10 to perform the actual deployment using rolls of polyimide film which we have proposed as a backbone for antenna arrays (Section 4.2). The three arms were deployed in a "Y" pattern, which is one possible configuration for a future lunar radio telescope array similar to that used by the NRAO Jansky Very Large Array [20].

Finally, on August 20, 2013 (Session 3), Astronaut Karen Nyberg remotely operated K10 to document the deployed telescope array (Figure 3, bottom). The primary objective of this final phase was to acquire high-resolution images of each antenna arm. These images serve two purposes: (1) in-situ, "as built" document of the deployed array; and (2) source data for locating and analyzing potential flaws (tears, kinks, etc.) that may have occurred during deployment.

Our data analysis [21] indicates that command sequencing with interactive monitoring is an effective strategy for crew-centric surface telerobotics: (1) planetary rover autonomy (especially safeguarded driving) enabled the human-robot team to perform missions safely; (2) the crew maintained good situation awareness with low effort using interactive 3-D visualization of robot state and activity; and (3) rover utilization was consistently in excess of 50% time; and (4) 100% of crew interventions were successful. In addition, we observed that crew workload was consistently low, which suggests that multi-tasking may be possible during telerobotic operations. A detailed description of the data collection, data analysis, and results is contained in [21].

We found that supervisory control is a highly effective strategy for crew-centric surface telerobotics. Subjective measurements made with the Bedford Workload Scale (BWS) [22] indicate that the task load was low. The BWS is a ten-point interval rating scale, which is based on the concept of "spare capacity" and which is encoded as a decision tree chart. The BWS provides subjective ratings of workload during (or immediately following) task performance. We presented the BWS chart to crew at random times throughout each session on a secondary laptop and recorded their subjective workload rating. During Session 1, workload varied on the BWS scale between 2 (low) and 3 (spare capacity for all desired additional tasks). In Session 2, workload was consistently and continuously 2 (low). Finally, during Session 3, workload ranged from 1 (insignificant) to 2 (low).

Using "Situation Awareness Global Assessment Technique" (SAGAT) [23] questionnaires, we determined that all three astronauts were able to maintain a high level of situation awareness (SA) during operations. We presented SAGAT questions to the crew at the same time as the BWS chart and used their responses to measure SA. In particular, we found that each astronaut was able to maintain all three SA levels (perception, comprehension, and projection) more than 67% of the time. From post-test debriefs, we also determined that interactive 3-D visualization of robot state and activity employed in the operator interface was a key contributing factor to achieving high levels of SA. Additionally, we observed that the increasing operational difficulty from Session 1 to Session 3 correlated with a decrease in SA.

The 2013 tests suggest that for future missions where astronauts would operate surface robots from an Earth-Moon L2 halo orbit or NRHO, it is important to design the system and operational protocols to work well with variable quality communications (data rates, latency, availability, etc.; see Section 3.2). In addition, for deep-space missions, it will be important to understand how efficiently and effectively a small crew of astronauts can work when operating robots largely independent of mission control support.

ESA has subsequently performed low latency telerobotics experiments aboard the ISS via Project METERON (Multipurpose End-to-End Robotic Operation Network) [24][25]. The goals of METERON include "technology validation, experiments in supervised autonomy, human-robot interaction in constrained environments, and inter-operability of groups of robots and control devices".

Future surface telerobotics testing with the ISS could be designed to test different mission objectives, such as field geology or sample collection [26]. The ISS presents a highly configurable and unique opportunity to study deep-space human mission concepts, including a high-fidelity spacecraft environment (microgravity, long-duration in space, in-flight stress, etc.) for crew. Potential benefits to future missions include: creating optimized crew training techniques and procedures, reducing operational risk and technology gaps, defining preliminary mission requirements, and estimating development and mission cost.

## 3.2. The Impact of Operational Constraints on Surface Telerobotic Exploration Efficiency

There are several operational constraints associated with using low latency telerobotics for deep space exploration as described above. These include increased communication latency, human control interfaces, camera placement, video conditions limited by available bandwidth, etc. Since astronauts will primarily use video feedback from surface assets to gain situation awareness, it is important to identify the operational constraints associated with video feedback. The bandwidth available between an orbiting control platform and ground assets will be variable. This necessitates quantifying operational video limitations, such as the threshold video frame rate, resolution, and colorscale. Quantifying these operational video limitations is important to ensure the success of low latency telerobotic exploration.

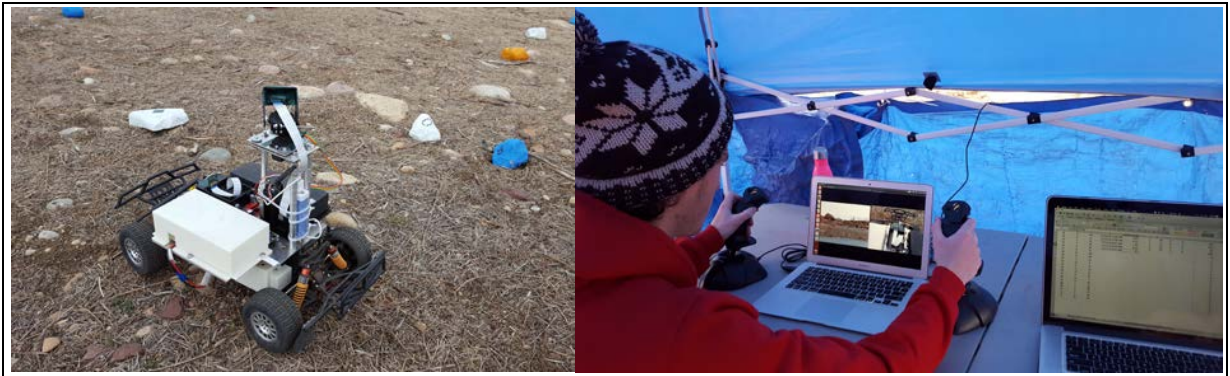

**Figure 4.** *Left:* The University of Colorado student-built, teleoperated rover and exploration targets (painted stones). *Right:* Operator controlling the rover joysticks ensuring exploration through the rover's twin cameras.

Over the past several years, a student-led project at the University of Colorado was undertaken to investigate the effects of video frame rate on an operator's ability to explore an unfamiliar environment using low latency telerobotics [27]. Frame rate was the only operational video constraint used in our experiment to simplify the experimental design. Previous research indicated that video frame rate is more important than video resolution and colorscale for successful situation awareness [28]. Our hypothesis was that a frame rate threshold exists such that once it is met, exploration efficiency is reduced to a point where operations are no longer effective. Such data will be helpful in the design of the telerobotics systems for the LOP-G. In order to test our hypothesis, we devised an experiment utilizing "interesting" exploration target objects, telerobotic operators, and a Telerobotic Simulation System (TSS). The target objects used were painted rocks with various symbols. The TSS consisted of a human operated rover controlled with joysticks via a radio-frequency transmitter/receiver and a suite of software used to adjust the video stream conditions. Pictures illustrating this experiment are shown in Figure 4.

The experiment took place on the University of Colorado campus within a crater-like landscape. Each trial had the operator search for a particular target object, a blue rock with an 'X' for example. Each trial had three possible frame rates: 4, 5, or 6 frames/sec. We used these frame rates after a precursor experiment, which tested frame rates ranging from 4 to 16 frames/sec, indicated that the threshold frame rate was most likely between 4 and 6 frames/sec. The frame rate for each trial was randomly distributed as well as the target object the operator had to find. In addition to recording the time to discovery for each trial, we also documented the number of times the rover became immobilized.

The data analysis began by examining the distribution of times to discovery for the exploration targets. We found this distribution to be inconsistent with a normal distribution at the 95% confidence level. This, then, determined the approach for studying the variance of each frame rate and subsequent post-hoc analyses. Next, we evaluated the mean absolute deviation from the median (ADM) or, effectively, the variance in times to discovery. Considering that our data do not fit a Gaussian distribution, an Analysis of Variance (ANOVA) was performed on the ADM. It was determined that the mean ADM across each frame rate was not equivalent. We then employed a post-hoc analysis to compare the different frame rate groups. We found that 5 and 6 frames/sec were indistinguishable from each other while 4 frames/sec had a significantly higher mean ADM with 95% certainty.

We then analyzed the mean time to discovery (MTD) with respect to frame rates as shown in Figure 5. We ran an ANOVA and found that at least one frame rate was statistically different from the others. Again, we used a post-hoc analysis to determine which frame rates were different from one another with respect to the MTD. The 5 and 6 frames/sec were found to be indistinguishable from each other while 4 frames/sec had a significantly higher MTD with 95% certainty. Thus, we discovered for the parameters of this particular experiment that a threshold of

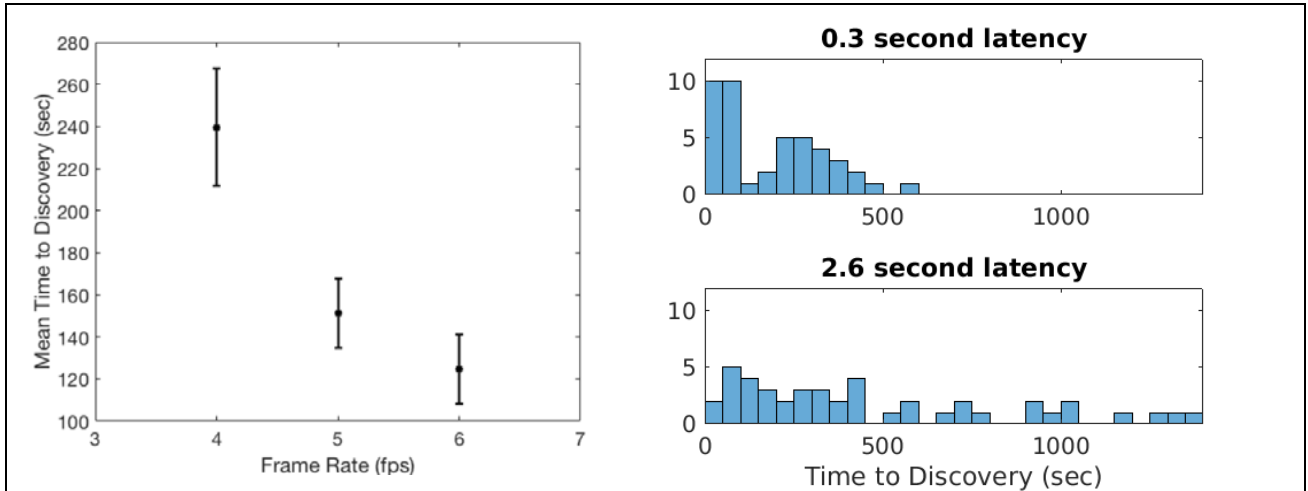

**Figure 5.** *Left:* The mean time to discovery (MTD) of target objects versus frame rate for the Colorado teleoperated rover shows a significant degradation, at the 95% confidence level, below 5 frames/second. *Right:* The dispersion of time to discovery increases abruptly with the 2.6 second latency condition. Trials at 2.6 seconds of latency show a 50% increase in the time to discovery versus a 0.3 second delay.

5 frames/sec exists, below which the operational effectiveness for exploration/discovery drops significantly. This result is consistent with video game experiments and other telerobotics experiments [29].

We followed our bandwidth experiment with an investigation on the effect of increased latency conditions [27]. Specifically, we examined how 2.6 seconds of latency affects the MTD of the target objects. 2.6 seconds was chosen because it is the best-case, round-trip communication time from Earth to a communication satellite at Earth-Moon L2 to the lunar far side. This experiment was conducted in the same manner as the previous with a slight variation. The frames/second was fixed at 5 and the introduced latency for each trial was randomly selected as either 0.3 seconds or 2.6 seconds. The rest of the experiment was conducted with the same rover and a similar course setup.

As with the bandwidth experiment, the data analysis began by examining the distribution of times to discovery for exploration targets. Again, we found the distribution to be inconsistent with a normal distribution at the 95% confidence level. We once again analyzed the MTD at each latency condition by performing an ANOVA. The two MTDs were statistically different from one another, in particular the MTD at 2.6 seconds of latency was significantly higher. There was a 150% increase moving from 0.3 seconds to 2.6 seconds latency as can be seen in Figure 5. Overall, the small amount of latency introduced produced a drastic worsening in real-time telerobotic exploration given the parameters of our experiment. It is important to note that the results of our experiment only hold for surface telerobotic exploration tasks with similar parameters used in our experiment; this includes rover speed, video colorscale, video resolution, etc. A study should be performed to investigate the interplay between these variables, such as how frame rate and rover speed interact with each other to affect telerobotic operations.

## 4. SCIENCE FROM THE LUNAR FAR SIDE

There are a number of exciting and important science goals to be pursued via a return to the Moon, especially on the unexplored lunar far side, as outlined in the NASA LEAG Roadmap [30]. In this section, we describe two examples that might be pursued early via low latency surface telerobotics from the Gateway. These are attractive examples because they address high priority goals from the U.S. National Research Council (NRC) studies and NASA roadmaps.

## 4.1. Human-Assisted Sample Return

If a human-assisted telerobotic sample return mission is developed, where might it land? A global lunar landing site study [31] investigated sites suitable for NRC-2007 [4] science and exploration objectives and determined the prime landing site for sample return missions is the Schrödinger impact basin within the South Pole-Aitken impact basin. As introduced in Section 2, that is exactly the type of mission that a Gateway, providing access and communication to a far side landing site, can facilitate.

The highest priority objective of the NRC-2007 report is to test the lunar cataclysm hypothesis, which suggests the Moon (and, thus, Earth) was severely bombarded by asteroids and comets circa 4 billion years ago. Because that event reshaped the surface of the Moon, may have resurfaced all of the terrestrial planets, and has been implicated in the early evolution of life, it is important to determine the duration of that bombardment. The Schrödinger basin (Figure 6) is the second youngest impact basin on the Moon, while the South Pole-Aitken (SPA) basin is the oldest impact basin on the Moon, so collecting samples produced by both events would bracket the duration of the hypothesized impact cataclysm.

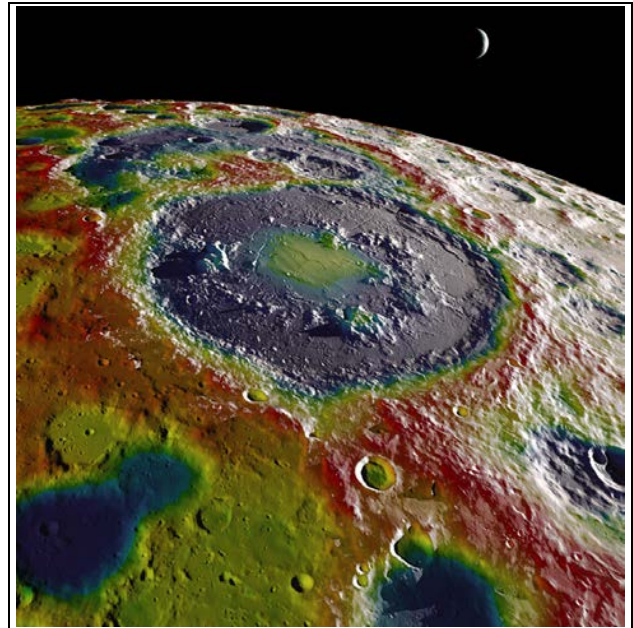

**Figure 6.** Orbital perspective of the ~320 km diameter Schrödinger basin on the lunar far side, looking from the north towards the south pole, with the Earth visible beyond the pole. GRAIL free-air gravity [35] mapped over LOLA- and LRO-derived terrain. NASA's Scientific Visualization Studio.

Moreover, because the Schrödinger basin is young, it will provide an opportunity for geologists to collect and analyze samples that have not been altered by later basin-forming events, which is one of the complexities that has confounded interpretations of Apollo samples from older, more complex sampling sites. Once the impact samples of Schrödinger are understood, it will then be possible to re-interpret with greater confidence the Apollo samples or any future samples from similarly complex terrains.

The second highest priority objective of the NRC-2007 report is to determine the age of the SPA basin, to anchor the basin forming epoch, so one can address both the first and second highest priority objectives with samples collected within the Schrödinger basin (e.g., [32]).

The Schrödinger impact basin is ~320 km in diameter and has a ~150 km diameter mountainous peak ring rising up to 2.5 km above the basin floor. This peak ring is composed of mid- to lower crustal material uplifted >20 km to the surface by the impact event [33]. Thus, samples of that peak ring can be used to test the lunar magma ocean hypothesis, another major concept that emerged from the Apollo program. Testing the lunar cataclysm and lunar magma ocean hypotheses are but two of the exciting scientific issues that can be addressed within the Schrödinger basin.

On the exploration side, Schrödinger also hosts an immense pyroclastic vent that spewed, in a volatile-rich cloud, magmatic debris across a portion of the basin floor. This vent may be the largest indigenous source of volatiles in the south polar region of the Moon [34]. In anticipation of the in-situ resource (ISRU) potential of that vent, it was one of the principal exploration targets during the Exploration Systems Mission Directorate (ESMD) phase of the Lunar Reconnaissance Orbiter mission.

Sample collection to address the science objectives requires mobility (hence a teleoperated rover), rather than a static lander. In also requires acquisition of specific samples from specific sites. Thus, a rover must be able to approach sample sites and have the dexterity to collect suitable samples. Because the Schrödinger basin is such an attractive scientific and exploration target, it has been used for several mission concept studies to drive operational trade studies and define mission requirements for geologic studies. For example, a human-assisted lunar sample mission lasting a single sunlit period was devised for a robotic rover teleoperated by crew on the Orion vehicle [36]. Because the Schrödinger basin is such a large and diverse target, potentially several completely unique missions can be flown to the basin. Thus, a robotic sample return mission over a period of 3 years, involving crew in a Lunar Orbiting Platform-Gateway, has also been explored [11]. We refer readers to those previous studies [11][36] for additional details about rover traverses, rover instrument payloads, types of samples collected, and mass of samples collected.

The mission timelines used in those studies were generated by mission architects for ISECG-member space agencies based upon anticipated launch capabilities. Both the short-duration and long-duration missions could address science objectives outlined by [4], but a larger number of those objectives could be met with the longer mission. Some of those science objectives can be addressed elsewhere on the Moon, but they would require longer traverses and, in many cases, many more landed missions. A global landing site study [31] found that Schrödinger is attractive, in part, because so much science can be accomplished in a relatively small area.

It is anticipated that crew will eventually be deployed to the lunar surface. Indeed, current U.S. national space policy directs astronauts to the lunar surface for "long-term exploration and utilization." A design reference mission scenario, involving five landing sites (Malapert massif, South Pole, Schrödinger basin, Antoniadi Crater, and center of the South Pole-Aitken basin) has been developed [37]. In that scenario, Orion docks with the LOP-G, crew land on the lunar surface, conduct a surface mission in two pressurized rovers, and then return to the Gateway. From the Gateway, crew return to Earth in Orion. The two rovers left on the surface would be teleoperationally driven to the next landing site using the communication relay provided by the LOP-G [38]. Thus, teleoperations are an essential component of both the robotic precursor missions and the subsequent human missions to the lunar surface.

### 4.2 Low Radio Frequency Observations of the Universe's Cosmic Dawn

A teleoperated rover can also deploy, construct, and assemble a low frequency radio telescope [1]. Radio observations below 100 MHz uniquely probe the earliest generations of stars in the Universe (Cosmic Dawn) [39]. The lunar far side affords a matchless platform for such observations, because this frequency range is contaminated on Earth by human-made signals (e.g., FM radio, digital TV) and is highly distorted by the Earth's ionosphere [40]. Neutral hydrogen fills the intergalactic medium surrounding the first stellar populations, and it is modified by heating and ionization as these first luminous objects begin to radiate. A hyperfine transition of neutral hydrogen emits photons (at rest wavelength 21-cm and frequency 1.42 GHz) with redshifted frequencies of 10-100 MHz which allows us to indirectly examine the birth and evolution of primordial stars for the first time.

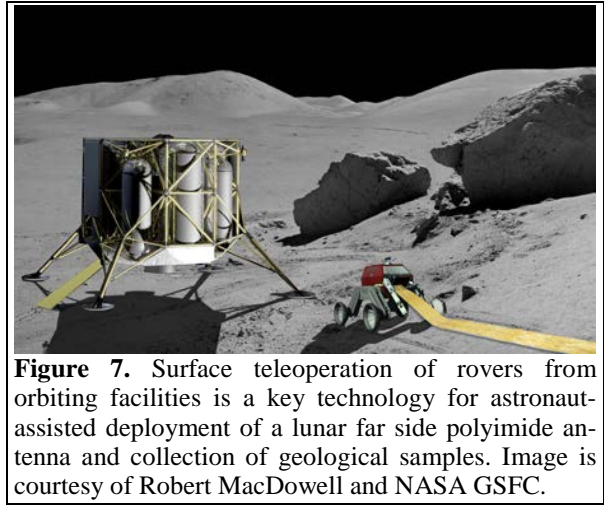

**Figure 7.** Surface teleoperation of rovers from orbiting facilities is a key technology for astronaut-assisted deployment of a lunar far side polyimide antenna and collection of geological samples. Image is courtesy of Robert MacDowell and NASA GSFC.

Two measurement approaches exist to survey the hydrogen around the first stars. First, the sky-averaged spectrum is the most basic quantity that can be measured with either a single antenna [39] or a compact array of dipole antennas [41]. Second, the power spectrum of hydrogen fluctuations allows the growth of structure (the "cosmic web") in the early Universe to be tracked. These observations require a more extensive array of 100's to 1000's of dipole antennas distributed over an area with a diameter of a few tens of kilometers. It is potentially more powerful than the sky-averaged spectrum, but it is also more difficult to measure because the signal from each structure is very small. Either measurement requires significant sensitivity within the radio-quiet environs of the lunar far side.

One viable approach to distribute an array of dipoles on the lunar surface begins with the deposition of electrically conducting antennas on a polyimide substrate [42]. The polyimide film with embedded antennas can then be rolled-out on the lunar surface with a teleoperated rover (Figure 7). This approach has significant benefits, most notably small volumes for transport and multiple deployment approaches. The gain/beam of the antenna will be dependent upon the sub-surface properties at the deployment site so this area will require further characterization. After deployment, the antennas can then be electrically phased for power spectrum measurements or incoherently combined for sky-averaged measurements. This later approach may be most appropriate for a pathfinder array requiring a smaller number of antennas but will be challenging due to the control of systematics. These trades are currently under study by our NASA Solar System Exploration Research Virtual Institute NESS (Network for Exploration & Space Science[1]) team.

The construction of a low radio frequency array will likely use a combination of supervised autonomy and telepresence. Initial deployment of the backbone polyimide substrate could likely be managed with moderate latency telerobotics directed from Earth after the terrain was mapped with a rover survey following an approach similar that discussed in the Section 3.1. Inspection and identification of flaws may require higher fidelity low latency telerobotics from the Gateway. Most importantly, further assembly of the array to connect transmission lines, power, electronics, and communication likely requires astronauts to manage in real-time. This aspect of the precision array assembly is most analogous to robot-assisted telepresence surgery where operators must be within their cognitive horizon (<0.5 second latency) relative to the patient [43]. The value of real-time astronaut-in-the-loop was demonstrated during several repair/replacement missions to the Hubble Space Telescope. A >50% reduction in precision, as suggested by our experiments described in Section 3.2, in going from teleoperation via the LOP-G to that of the best case from Earth will likely be inadequate to assemble a radio array on the lunar surface. We have begun new laboratory experiments of telescope array assembly to test the effects of latency on precision construction tasks.

## 5.    SUMMARY AND CONCLUSIONS

In this paper, we described the synergy between the Orion crew vehicle and the Lunar Orbiting Platform-Gateway that enables support of low latency surface telerobotics on the Moon. Crew are anticipated to be present at least 30 days per year at the Gateway. During this time, operational efficiencies are expected to allow significant crew time dedicated to science experiments involving surface telerobotics. When the Gateway is unoccupied, it will function like other robotic planetary spacecraft with remote control from Earth. Importantly, the LOP-G will provide communication to and telemetry from surface assets, especially from the lunar far side. In addition, the Gateway could support other mission services including remote sensing, high performance computation and data storage to reduce the load on surface rovers/experiments, as well as providing positioning and timing data.

To prepare for the first low latency surface teleoperation of robots on the Moon, we have performed several precursor experiments. First, we conducted a simulation of a Gateway teleoperated lunar rover using the International Space Station as a surrogate for the LOP-G and the K10 rover in an analogue lunar terrain. Three crew sessions were devoted to studying site surveying, radio telescope deployment, and inspection of the radio telescope. The experiment was successful in demonstrating an effective strategy for crew-centric surface telerobotics. The ISS crew maintained good situation awareness with low effort using the interactive tools provided by our team. Second, we conducted a student-led experiment to probe the operational constraints associated with low latency surface telerobotics, testing the effects of video bandwidth and latency. The metric used to investigate these effects was the mean time to discovery (MTD) of exploration targets. We found that the MTD dropped significantly, at the 95% confidence level, when the video frame rate (i.e., bandwidth) dropped below 5 frames/sec. We also discovered that the MTD increased by 150% in moving from a latency of 0.3 seconds (equivalent of round-trip between the LOP-G and lunar surface) compared to a latency of 2.6 seconds (best possible light speed round trip travel time from Earth through the Gateway and then to the lunar far side). These results will figure into the planning for science experiments on the lunar far side.

Two examples of science from the lunar far side facilitated by low latency surface telerobotics were presented that correspond to goals from NRC Decadal Surveys. First, we described a human-assisted sample return mission focused on the Schrödinger impact basin. Samples returned from this region of the South Pole-Aitken Basin will

---

provide a key test of the lunar cataclysm hypothesis which suggested that the Moon and Earth suffered a late heavy bombardment period about 4 billion years ago. Sample collection to address the science objectives of this mission require the mobility provided by a teleoperated rover. This mission requires specific samples at specific sites, with a teleoperated rover having the flexibility to collect suitable samples. After the crew departs the Gateway, the rover could continue lower fidelity operations supervised from Earth. Second, a teleoperated rover can also deploy, construct, and assemble a low radio frequency telescope on the radio-quiet lunar far side. This telescope will have the unique capability of probing the epoch of formation of the first stars and galaxies, as well as investigate the possible interaction of dark matter with neutral hydrogen at frequencies (redshifts) down to 10 MHz (z=140) [44]. The construction of an array of low radio frequency telescopes will likely use a combination of supervised autonomy and telerobotics. Assembly to connect power, electronics, and communications will likely require the precision of low latency teleoperation.

In addition to the early science that will be accomplished using low latency surface telerobotics during extended stays at the Gateway, key technologies and operational strategies will be matured that will likely feed-forward to the first human missions to Mars.

## ACKNOWLEDGEMENTS

This work was directly supported by the NASA Solar System Exploration Virtual Institute cooperative agreements 80ARC017M0006 to J. Burns (PI) and NNA14AB07A to D. Kring (PI). The Colorado student telerobotics project was funded in part by the Lockheed Martin Space Systems Company. The 2013 ISS Surface Telerobotics project was supported by the NASA Space Technology Mission Directorate. We thank Robert MacDowall, Joseph Lazio, Dayton Jones, Raul Monsalve, Chris Norman, and Wendy Bailey for their contributions to components of the projects discussed herein. We appreciate the constructive comments of three reviewers which have improved this paper.